\documentclass[letterpaper]{jpconf}
\usepackage{iopams}

\newcommand{\defeq}{:=}
\newcommand{\R}{\mathbb{R}}
\newcommand{\C}{\mathbb{C}}
\newcommand{\tens}{\otimes}   
\newcommand{\cH}{\mathcal{H}}

\newcommand{\cS}{\mathcal{S}}
\newcommand{\cA}{\mathcal{A}}
\newcommand{\cB}{\mathcal{B}}
\newcommand{\cC}{\mathcal{C}}

\begin{document}

\title{Probabilities in the general boundary formulation}
\author{Robert Oeckl}
\address{Instituto de Matem\'aticas, UNAM, Campus Morelia,\\
C.P.~58190, Morelia, Michoac\'an, Mexico\\
$[$version with corrections$]$}
\ead{robert@matmor.unam.mx}



\begin{abstract}
We give an introductory account of the general boundary formulation of
quantum theory. We refine its probability interpretation and emphasize
a conceptual and historical perspective. We give motivations from
quantum gravity and illustrate them with a 
scenario for describing gravitons in quantum gravity.

\end{abstract}


\section{A geometric-algebraic approach to quantum theory}

Since we want to contrast the general boundary formulation with the
standard formulation of quantum theory, we should state precisely what
we mean by the latter. Firstly, there is a \emph{time}
variable $t\in\R$ which is classical and provided from the
outset. Secondly, there is a Hilbert space $\cH$ of \emph{states}. In the
Schr\"odinger picture, which we will use, a state is thought to encode
the physical
condition of the system under consideration at a given time. Thirdly,
a state evolves in time according to a dynamical law. The time
evolution from time $t_1$ to time $t_2$ is encoded in an operator
$U(t_1,t_2):\cH\to\cH$. For consistency we must have
$U(t_2,t_3) U(t_1,t_2)=U(t_1,t_3)$ for any $t_1<t_2<t_3$. We will
refer to this as the \emph{composition rule}. Note that we may
describe $U(t_1,t_2)$ through its matrix elements, i.e.,
\emph{transition amplitudes}. We write the transition amplitude from
an initial state $\psi$ at time $t_1$ to a final state $\eta$ at time
$t_2$ as usual as $\langle \eta | U(t_1,t_2)|\psi\rangle$.
Fourthly, the modulus square of a transition amplitude encodes a
probability in a measurement process: $|\langle \eta |
U(t_1,t_2)|\psi\rangle|^2$ encodes the probability of measuring the
state $\eta $ at time $t_2$ given that the state $\psi$ was prepared
at time $t_1$. Conservation of probability in time requires that the
operators $U(t_1,t_2)$ preserve the inner product, i.e., be
\emph{unitary}.

Of course, actual quantum theories encompasses more than what we have
just described. However, it seems fair to say that realistic theories of
quantum mechanics and quantum field theory contain at least the
elements enumerated above.\footnote{\label{fn1}This characterization
  might not
apply to certain frameworks for quantum field theory in curved
spacetime, even for a fixed foliation. However, there is as yet no
experimental test of quantum field theory in curved spacetime which
could confirm such a framework.} Hence, we shall refer to these
elements as defining the \emph{standard formulation of quantum theory}.

To prepare the ground for the general boundary formulation let us
think of the standard formulation as associating algebraic to geometric
structures as follows. The geometric structures are points in time
and intervals of time. The algebraic
structures are states and transition amplitudes.
To each point $t\in\R$ in time we associate a Hilbert space $\cH_t$ of
states. These state spaces are simply copies of the usual state
space $\cH$ and the
labeling by a time is only a formality at this point.
To each time interval $[t_1,t_2]\subset\R$ we associate a
linear map
$\rho_{[t_1,t_2]}:\cH_{t_1}\tens\cH_{t_2}\to\C$, called \emph{amplitude
map}.\footnote{The second tensor factor in the domain of
  $\rho_{[t_1,t_2]}$ should really be the dual space $\cH_{t_2}^*$ of
    $\cH_{t_2}$. This is because it represents a ``bra''-vector in
    contrast to the first argument which represents a
    ``ket''-vector. For simplicity of presentation, we gloss over this
  fact here as well as in the following.}
This sends a pair of an initial
and a final state
$(\psi,\eta)$ to the transition amplitude
$\rho_{[t_1,t_2]}(\psi\tens\eta)=\langle
\eta|U(t_1,t_2)|\psi\rangle$. The amplitude map
may be thought of as a map from the
Hilbert space associated to the boundary $\partial
[t_1,t_2]=t_1\cup t_2$ of the time interval to the complex numbers.
More generally, associate to a union of points in time
$t_1\cup\cdots\cup t_n$ the product Hilbert space $\cH_{t_1\cup\cdots\cup
  t_n}\defeq\cH_{t_1}\tens\cdots\tens\cH_{t_n}$.
The composition rule may be
seen as establishing a correspondence between geometric and algebraic
operations: The geometric gluing of time intervals is put into
correspondence with the algebraic composition of transition
amplitudes. Concretely, the gluing $[t_1,t_3]=[t_1,t_2]\cup [t_2,t_3]$
is put into correspondence with the composition
$\rho_{[t_1,t_3]}(\psi\tens\eta)
=\sum_i\rho_{[t_1,t_2]}(\psi\tens\xi_i)\rho_{[t_2,t_3]}(\xi_i\tens\eta)$,
where $\{\xi_i\}$ denotes an ON-basis of $\cH_{t_2}$.

The geometric-algebraic viewpoint becomes more compelling when we
consider quantum field theory in (possibly curved) spacetime.
Now the analogues of the points in
time are spacelike hypersurfaces in spacetime. Again we have a state
space $\cH_\Sigma$ associated with each spacelike hypersurface
$\Sigma$.
The analogues of the time-intervals are now spacetime regions bounded
by two such hypersurfaces (an initial and a final one). Again we
associate a linear amplitude map $\rho_M:\cH_{\partial M} \to\C$
to such a region $M$ with boundary $\partial M=\Sigma\cup\Sigma'$ and
$\cH_{\partial M}=\cH_{\Sigma}\tens\cH_{\Sigma'}$. As before, this map
sends a pair of initial and final state to the associated transition
amplitude. And again, we have a composition rule associating to a
gluing of spacetime regions the composition of the corresponding
amplitude maps.

Let us go one step
further: Associate state spaces to general hypersurfaces and
amplitude maps to general spacetime regions.
An immediate observation is that now different hypersurfaces might
have not only different geometry, but even different topology. It is
thus to be expected that the associated state spaces are not
(naturally) isomorphic. Hence the previously formal distinction
between them becomes a necessity. In this sense \emph{the} state-space
of a theory no longer exists.
Another observation is that the
boundary of a general spacetime region does not necessarily decompose
into an ``initial'' and ``final'' hypersurface. Only in special cases
it is possible to interpret the amplitude
map $\rho_M:\cH_{\partial M}\to\C$ for a region $M$ with boundary
$\partial M$ as a transition amplitude between
state spaces associated with boundary components.
The composition rule can be generalized to this context as well as the
notion of unitarity.

What we have described so far is the structural part of the \emph{general
boundary formulation of quantum theory}.
We turn to its
interpretational part later.
See \cite{Oe:gbqft} for a more comprehensive
account and
\cite{Oe:2dqym} for a refined version of the system of rules or
\emph{axioms} involved.

\section{Historical Remarks}

\subsection{Dirac's attempt}

Remarkably, the idea of associating amplitudes to general spacetime
regions already appears in a paper by Dirac
\cite{Dir:lagqm}.
Dirac introduces what he calls
``generalized transformation functions'', which are essentially the same
as the generalized amplitudes considered here. What is more, he even
proposes the spacetime gluing rule. However, his idea seems to have
not been pursued at the time.\footnote{Indeed, parts of this paper
appear as well in
Dirac's monograph on quantum mechanics \cite{Dir:qm}. However, the part on
``generalized transformation functions'' is missing there.}
This is not surprising, for several reasons:
\begin{enumerate}
\item \label{e1} The lack of need for such a formulation.
\item \label{e2} The technical difficulties in realizing the proposal.
\item \label{e3} The lack of a physical interpretation of the generalized
  amplitudes.
\end{enumerate}

Let us elaborate on these reasons in turn.
Concerning point (\ref{e1}), it turned out that
non-relativistic quantum mechanics as well as quantum
field theory get along very well with the usual type of transition
amplitudes. Indeed, it is only in quantum gravity that one should
expect this to no longer to be the case.
In this sense Dirac's proposal was certainly ahead of its time.
We will elaborate on this point later.

Concerning point (\ref{e2}), standard quantization prescriptions
rely on initial value problems. Since the
differential equations of the classical theory are generally
hyperbolic, this is normally ensured by considering the space of
configuration data and its first derivatives on spacelike
hypersurfaces. More general hypersurfaces do not admit a
correspondence between classical solutions and boundary data of such a
simple form.
Furthermore,
most quantization prescriptions describe time-evolution in an
infinitesimal fashion. Exponentiation then leads to a spacelike
hypersurface sweeping out a region of spacetime of interest, yielding
the associated time-evolution operator.
There is
no analogous way to describe a general spacetime region and its
associated amplitude.
These two problems require a considerable
modification of quantization prescriptions. This is probably
the chief technical difficulty of the general boundary
formulation.

Concerning point (\ref{e3}), it
is not clear how a general amplitude should give rise to
a probability.
Remarkably, Dirac himself, in the last paragraph of \cite{Dir:lagqm},
attempted a probability interpretation, which unfortunately seems to be
untenable.
What is more, one might think that crucial consistency
properties such as probability conservation require a temporal
ordering of states, hence rendering any attempt at a consistent
probability interpretation futile. While this is probably the most
important a priori objection to the general boundary formulation it
turns out to be unfounded. We will come to the probability interpretation
in the next section.

\subsection{Topological quantum field theory}

A mathematical abstraction of the framework we have described so far
is known as \emph{topological quantum field theory}, see e.g.\
\cite{Ati:tqft}. This arose in the 1980s and was strongly inspired by
the path integral approach to quantum field theory. The latter goes
back to Feynman's seminal paper \cite{Fey:stnrqm} which in turn was
inspired by Dirac's \cite{Dir:lagqm}. However, Feynman himself did not
write anything about the generalized amplitudes.

Since topological quantum field theory is a mathematical framework the
three reasons mentioned above for the lack of success of Dirac's idea
do not apply to it. Concerning point (\ref{e1}), there are indeed
interesting
applications. Most notably, topological quantum field theory has given
rise to a whole new branch of algebraic
topology including the discovery of new invariants of knots and of
3-manifolds. While it is also applied to models in mathematical
physics, these are toy models or auxiliary models where the technical
difficulties
referred to in point (\ref{e2}) do not occur. These models
also do not possess or require a direct interpretation in the sense
of point (\ref{e3}).

\section{Probability Interpretation}

We turn now to the interpretational part of the general
boundary formulation. In the
center stands the interpretation of generalized amplitudes as
giving rise to physical probabilities.
Probabilities in quantum theory are generally
\emph{conditional probabilities}. More specifically, such a
probability usually depends on two type of data: Data that describes
\emph{knowledge} or \emph{preparation} and data that describes a
\emph{question} or \emph{observation}. In the simplest case we are
looking for the probability to observe a specific state given that
some other specific state was prepared.

In the general boundary formulation the dependence of
probabilities on these two types of data is preserved. Consider a
process taking place in a spacetime region $M$ with boundary
$\partial M$. Then, both type of data are encoded through closed
subspaces of the state space $\cH_{\partial M}$. Let
  $\cS\subset\cH_{\partial M}$ represent \emph{preparation} or
  \emph{knowledge}, and
  $\cA\subset\cH_{\partial M}$ represent \emph{observation} or
  the \emph{question}.
The probability that the system is described by $\cA$ given
that it is described by $\cS$ is:
\begin{equation}
   P(\cA|\cS)=\frac{|\rho_M\circ P_\cS\circ P_\cA|^2}
   {|\rho_M\circ P_\cS|^2} .
\label{eq:prob}
\end{equation}
Here, $P_\cS$ and $P_\cA$ are the orthogonal projectors onto the
respective subspaces and $\circ$ represents composition of
maps. Hence, the expressions in numerator and denominator
that the norm square is taken of are linear maps $\cH_{\partial
  M}\to\C$. The norm of such a map is defined here as
follows.\footnote{There are a few subtleties that we are not detailing
  here. In particular, $\rho_M$ is generically not bounded. Thus,
  $\cS$ must be ``small
  enough'' such that $\rho_M\circ P_\cS$ is bounded. This condition is
  satisfied in standard situations.}
Let $\alpha:\cH_{\partial M}\to\C$ be a bounded linear map. Then there
exists
  $\xi\in\cH_{\partial M}$ such that $\alpha(\psi)=\langle \xi
  ,\psi\rangle\quad\forall\psi\in\cH_{\partial M}$.
Define $|\alpha|:=|\xi|$.

$P(\cA|\cS)$ has the properties expected
of a (quantum mechanical) probability:
\begin{itemize}
\item
By construction $P(\cA|\cS)$ takes values in the interval
$[0,1]$.\footnote{It might happen that the denominator is zero. In
this case the numerator is also zero and $P(\cA|\cS)$ is
undefined. Physically this means that the knowledge encoded in $\cS$
does not correspond to any allowed process.}
\item
 Given two
mutually exclusive observations encoded by orthogonal subspaces
$\cA_1$ and $\cA_2$ the respective probabilities are additive, i.e.,
$P(\cA_1\oplus\cA_2|\cS)=P(\cA_1|\cS)+P(\cA_2|\cS)$.
\item
 The
probability for an arbitrary outcome is $P(\cH_{\partial M}|\cS)=1$
for any $\cS$.
\item
 If $\cA\subseteq\cB\subseteq\cC\subset\cH_{\partial M}$
(read: $\cA$ implies $\cB$ implies $\cC)$ we have the chain rule
$P(\cA|\cC)=P(\cA|\cB)P(\cB|\cC)$.
\end{itemize}

To see how the expression (\ref{eq:prob}) reduces to a standard
transition probability, consider a region that is given by a time
interval $[t,t']$. Thus, $\cH_{[t,t']}=\cH_t\tens\cH_{t'}$ and
$\rho_M(\psi\tens\eta)=\langle \eta | U(t'-t) | \psi\rangle$
as explained
above. We want to calculate the probability that $\eta$ is observed
given that $\psi$ was prepared. The preparation corresponds to the
subspace $\cS=\psi\tens\cH_{t'}\subset\cH_{[t,t']}$ while the
observation corresponds to the subspace
$\cA=\cH_t\tens\eta\subset\cH_{[t,t']}$. It then turns out that formula
(\ref{eq:prob}) yields
$P(\cA|\cS)=|\langle \eta | U(t'-t) | \psi\rangle|^2$ as required. For
more complex examples of how (\ref{eq:prob}) recovers the correct
probabilities of standard quantum theory see \cite{Oe:gbqft}.

Note a slight difference to the previous presentation of
the probability interpretation in \cite{Oe:gbqft}. There, the subspace
$\cA$ was restricted to be a subspace of $\cS$ as well. Lifting this
restriction represents more a formal than a physical
difference. Conceptually, making $\cA$ a subspace of $\cS$ just means
taking into account the knowledge about the measurement when the
question is asked. In particular, if $P_\cS$ and $P_\cA$ commute we
can replace $\cA$ by $\cA\cap\cS$ without any change to $P(\cA|\cS)$.

A property that is central to the consistency of the probability
interpretation of quantum theory is \emph{probability
  conservation}. Usually, this refers to conservation of probability
\emph{in time}. 
The present probability interpretation allows to extend this to a
more general notion of probability conservation \emph{in spacetime}.
Consider a spacetime region $M$ and an adjacent region $N$ that we can
think of as ``deforming'' $M$ to $M'=M\cup N$.\footnote{A technical
  definition
  of ``deformation'' could be: $N$ should be contractible onto
  (a part of) the boundary of $M$.}
The amplitude map $\rho_N:\cH_{\partial N}\to\C$ associated with $N$
induces a map $\tilde{\rho}:\cH_{\partial M}\to\cH_{\partial M'}$. Now
let $\cS\subset\cH_M$ and $\cA\subset\cH_M$ be subspaces determining a
measurement in the sense discussed above. Define the subspaces
$\cS'\defeq\tilde{\rho}(\cS)\subset\cH_{M'}$ and
$\cA'\defeq\tilde{\rho}(\cA)\subset\cH_{M'}$. Then
$P(\cA|\cS)=P(\cA'|\cS')$, i.e.,
the probability
for observing $\cA$ given $\cS$ on $\partial M$ is the same
as that for observing $\cA'$ given $\cS'$ on $\partial
M'$. Probability is conserved for ``evolution'' through the spacetime
region $N$. For a more detailed discussion of this example, see
\cite{Oe:gbqft}.
If $M$ is a time interval $[t_1,t_2]$ and $N$ an adjacent time
interval $[t_2,t_3]$ we recover the standard notion of probability
conservation in time as a special case.

\section{The general boundary formulation as an extension of quantum
  theory}

The emphasis on spacetime might lead one to think of the
general boundary formulation as a generalization of quantum field
theory rather than quantum theory as such. However, quantum field
theory can be reduced to the standard formulation of quantum theory
while the general boundary formulation cannot.\footnote{But recall
  footnote~\ref{fn1}.} This is why we prefer to see it as
an extension of quantum theory itself.
On the other hand, if we take spacetime to be the real line of time,
we get back exactly the standard formulation of quantum theory (plus a
more general probability interpretation). In
this sense it is a special case of the general boundary
formulation. Insofar quantum field theory is also trivially
encompassed by the general boundary formulation. However, the
conjecture is that realistic quantum field theories can be extended to
quantum field theories based on the general boundary
formulation with spacetime being taken to be what it is, rather than
just a time axis. For examples and more details on the application to
quantum field theory, see
\cite{Oe:timelike,Oe:gbqft,Oe:kgtl,Oe:2dqym}.

It is clear already that the nature of the spacetime regions and
hypersurfaces considered must depend heavily on the theory under
consideration. As already discussed, in non-relativistic
quantum mechanics they would be intervals and points on the real line
(representing time). In standard quantum field theory they would be
pieces of Minkowski spacetime. (Interestingly, one can try to use such
a spacetime picture also in non-relativistic quantum mechanics, leading
to effects otherwise typical only of quantum field theory
\cite{Oe:boundary}.) In quantum field theory in curved spacetime they
would be pieces of spacetime with a general Lorentzian metric.

General relativity teaches us that the metric is a dynamical field
just like, say, the electromagnetic field. Hence, quantum gravity
should be defined on spacetime regions and hypersurfaces without
prescribed metric. This is sometimes referred to as \emph{background
independence}. Hence, spacetime regions and hypersurfaces should be
merely differentiable manifolds. Indeed, a chief virtue of the
general boundary formulation is that it admits such background
independent theories.\footnote{Sometimes ``background independence'' is
taken to mean also that there is no differentiable structure or not
even the structure of a (topological) manifold. The latter case would
not fit into the general boundary formulation, at least not in its
present from.}

\section{Towards quantum gravity}

\subsection{The quantum cosmology problem}

A main motivation for the general boundary formulation comes from
quantum gravity \cite{Oe:catandclock,CDORT:vacuum}. We
illustrate this with a well known problem that
we shall call the
\emph{quantum cosmology problem}.
If we look naively at standard quantum (field) theory
as a
description of our world, a state describes the whole universe at
a given time (or spacelike hypersurface). Of course, we are usually
interested in describing specific, localized systems and do not want
to bother with all the rest of the universe. Quantum field theory
allows us to do precisely that, telling us that distant systems can be
treated as independent. Underlying this are powerful properties such as
\emph{causality} and \emph{cluster decomposition}. In most
cases we can accurately describe a local system as if it was living
alone in an otherwise empty Minkowski universe.

In quantum gravity there is no metric background to separate
systems. What is worse, diffeomorphism gauge symmetry makes any kind
of (even relative) localization difficult. In particular, there is
no causality or cluster decomposition property from the outset. Hence,
at least a priori we cannot avoid that states are now really states of
the whole universe. Apart from technical problems this also prompts
deep conceptual problems such as to the meaning of quantum theory
without an outside observer. It seems we have to do \emph{quantum
  cosmology}.

While it might very well be that the mentioned problem can be solved
within the standard formulation of quantum theory, it can be
avoided in the general boundary formulation.
State spaces, amplitudes and probabilities referring to \emph{local}
regions of spacetime allow to describe their physics independent of the
physics outside. In particular, there is now no difficulty in placing the
observer outside of the quantum mechanical process under
consideration. Indeed, this suggests that we should \emph{only} allow
local regions in a quantum theory of gravity. Infinitely extended
regions or regions ``wrapping around the universe'' would not be
admissible.

\subsection{Graviton scattering: A scenario}

To understand more concretely how predictions in a
quantum theory of
gravity could be formulated we consider the following semiclassical
scenario. Consider a 4-ball shaped region $M=B^4$ in spacetime with
boundary $\partial M=S^3$. We suppose that there is a semiclassical
sector of the theory such that $M$ may be described as a piece of
Minkowski spacetime with small fluctuations. Formalizing this, the
state space associated with the boundary of $M$ contains a sector
$\cH_\mathrm{lin}$ describing this regime and
can be decomposed as $\cH_{\partial M}
=\cH_\mathrm{lin}\oplus\cH_\mathrm{nlin}$.

It should now be expected that $\cH_\mathrm{lin}$ is (approximately) a
Fock space with graviton states. What is more, learning from a related
quantum field theory example \cite{Oe:kgtl} we expect it to (approximately)
factorize into in- and out-states,
$\cH_\mathrm{lin}=\cH_\mathrm{in}\tens\cH_\mathrm{out}$. Now pick an
$n$-graviton state $\psi_{p_1,\cdots,p_n;\mathrm{in}}$ in
$\cH_\mathrm{in}$ and an $m$-graviton state
$\psi_{q_1,\cdots,q_m;\mathrm{out}}$ in $\cH_\mathrm{out}$. The
associated scattering probability (density) would be given by
$P(\cA|\cS)$ with 
$\cA=\psi_{p_1,\cdots,p_n;\mathrm{in}}\tens\cH_\mathrm{out}$ and
$\cS=\cH_\mathrm{in}\tens\psi_{q_1,\cdots,q_m;\mathrm{out}}$.
A few remarks are in order:
\begin{itemize}
\item Although the scenario discussed is that of a \emph{semiclassical}
  description, there is no approximation or reduction performed before
  quantization. All
  objects under consideration would be objects of the full quantum theory.
\item The factorization of the subspace $\cH_\mathrm{lin}$ into in-
  and out-states should not be expected to extend to a factorization
  of the full state space $\cH_{\partial M}$.
\item The detailed results will depend on how exactly we choose
    $\cH_\mathrm{lin}$
    in $\cH_{\partial M}$, in which way it approximates a Fock space,
    up to which energies, etc. One might conjecture that these
    ambiguities are related to the
    renormalization ambiguities of perturbative quantum gravity.
\end{itemize}

For first steps in applying (so far only the structural part of) the
general boundary formulation in a loop quantum gravity / spin foam context
see
\cite{CDORT:vacuum,Oe:bqgrav,Rov:qg,Rov:gravprop}.

\ack

I would like to thank Rafael Sorkin for pointing me to Dirac's
article \cite{Dir:lagqm}.

\bibliography{stdrefs}
\bibliographystyle{amsordx}

\end{document}